# Mineralogy and Surface Composition of Asteroids

**Vishnu Reddy**
Planetary Science Institute

**Tasha L. Dunn**
Colby College

**Cristina A. Thomas**
NASA Goddard Spaceflight Center

**Nicholas A. Moskovitz**
Lowell Observatory

**Thomas H. Burbine**
Mount Holyoke College

Methods to constrain the surface mineralogy of asteroids have seen considerable development during the last decade with advancement in laboratory spectral calibrations and validation of our interpretive methodologies by spacecraft rendezvous missions. This has enabled the accurate identification of several meteorite parent bodies in the main asteroid belt and helped constrain the mineral chemistries and abundances in ordinary chondrites and basaltic achondrites. With better quantification of spectral effects due to temperature, phase angle, and grain size, systematic discrepancies due to non-compositional factors can now be virtually eliminated for mafic silicate-bearing asteroids. Interpretation of spectrally featureless asteroids remains a challenge. This paper presents a review of all mineralogical interpretive tools currently in use and outlines procedures for their application.

## 1. INTRODUCTION

Physical characterization of small bodies through the use of reflectance spectroscopy provides insight into the diversity of chemical compositions in the Solar System and enables meteorites in our terrestrial collections to be linked to specific parent bodies in space. Since the publication of *Asteroids III* there have been numerous advances in the mineralogical characterization of asteroid surfaces using visible/near-infrared (VIS/NIR) and mid-IR spectra. One of the most significant developments in ground-based characterization of small bodies over the last decade is the commissioning of the SpeX instrument in the NASA Infrared Telescope Facility (IRTF) on Mauna Kea, Hawai'i. The ideal spectral resolution (~few hundred), wavelength range (0.8 –5 µm), and high altitude of Mauna Kea have facilitated the mineralogical characterization of asteroids down to a visual magnitude of 19.

In addition to ground-based observations, the last decade has been witness to a series of successful spacecraft missions, including flybys of (2867) Šteins, (21) Lutetia, (4179) Toutatis, a rendezvous mission to asteroid (4) Vesta, and a sample return from near-Earth asteroid (25143) Itokawa. These missions have helped study the shape, morphology, and surface composition of asteroids over a wide range of sizes, from Itokawa (535 m) to Vesta, which (at 525 km) is the second most massive object in the asteroid belt. Two more sample return missions are planned for the next decade, both of which are targeting primitive near-Earth asteroids. Hayabusa-2 calls for a sample return of C-type asteroid (162173) 1999 JU3, and OSIRIS-Rex will return a sample from the potentially hazardous asteroid Bennu.

Our ability to track meteorite falls back to their possible parent asteroids has benefited from increased interest in the monitoring of fireballs, the streak of light generated as a meteorite comes through the atmosphere. The Chelyabinsk meteorite, which exploded over Russia on February 15, 2013, is the most well-documented fireball in modern history. The combination of all sky camera data from networks such as Cameras for Allsky Meteor Surveillance, the Desert Fireball Network in Australia and NASA's All Sky Fireball Network, and weather radar data has helped in the recovery of several meteorites with accurate orbits in the last decade (Bland et al. 2009; Fries et al. 2014). Though a rare occurrence, the discovery of Earth impactors prior to impact, such as 2008 TC3, has enabled us to compare recovered meteorite samples to Earth-based telescopic observations (Jenniskins et al. 2009).

Advances in the collection and quality of spectral data have necessitated improvements in our ability to interpret these data. New formulas for deriving mineralogies from VIS/NIR spectra with prominent 1- and 2-µm absorption bands have been developed (Burbine et al. 2007; Dunn et al. 2010; Reddy et al. 2011), and the effects of non-compositional parameters have been well characterized, allowing for more accurate interpretations of asteroid surface mineralogies (e.g. Burbine et al. 2009; Sanchez et al. 2012, Reddy et al. 2012). For spectra devoid of prominent absorption features at visible and near-IR wavelengths, radiative transfer models can provide insight into mineral compositions and abundances. Another area of focus is the 3-µm region where hydrated silicates and organics have distinct absorption features. Much progress has also been made on spectroscopy of asteroids and meteorites at mid-IR wavelengths (5 – 25 µm).

In this chapter we review advances made over the last decade in determining mineralogy and surface composition of asteroids at wavelengths ranging from the visible to the mid-infrared. We focus on asteroids showing silicate absorption features as they have the best laboratory spectral calibrations to derive mineralogy with remote sensing data. We also examine the role that spacecraft have played in improving our understanding of asteroid surface morphology and composition. Lastly, we discuss the connections between meteorite groups and potential asteroid sources that have been made as a result of these advances in the collection and interpretation of asteroid spectra.

## 2. MINERALOGICAL CHARACTERIZATION

Visible/near-infrared (VIS/NIR) spectra (0.3-2.5µm) have been widely used to determine mafic mineral abundances and compositions of asteroids since the relationship between spectral properties (i.e. absorption features) and mineralogy in meteorites was first recognized (e.g. Adams 1975; Burns et al. 1972; Cloutis et al. 1986; Cloutis and Gaffey 1991). Most previous work examining the spectral properties of silicate minerals has focused on olivine and pyroxene, both of which have diagnostic spectral properties in the VIS/NIR.

Olivine and pyroxene are the primary mineral phases in most primitive meteorites and in many types of non-chondritic meteorites. The structure of olivine consists of individual silicon-oxygen tetrahedra linked by divalent Mg and Fe atoms that occupy the M1 and M2 structural sites. Substitution of Mg and Fe between the M1 and M2 sites results in a solid solution series defined by two compositional endmembers: the Fe-endmember fayalite ($Fe_2SiO_4$) and the Mg-endmember forsterite ($Mg_2SiO_4$). Pyroxenes are a group of chain silicate minerals that share the chemical formula (M2) (M1) $(Si, Al)_2 O_6$. Three pyroxene subgroups have been defined based on occupancy of the M2 site. In low-Ca pyroxenes, the M2 site is occupied by Fe or Mg, in high-Ca pyroxenes by Ca, and sodium pyroxenes by Na. Because high-Ca pyroxenes have monoclinic symmetries, they are often referred to as clinopyroxenes. The term orthopyroxenes is commonly used when referring to the orthorhombic low-Ca pyroxene endmembers, enstatite ($Mg_2Si_2O_6$) and ferrosilite ($Fe_2Si_2O_6$). For a more in depth discussion of the structure and composition of the olivine and pyroxene groups, we refer the reader to Deer et al. (1966).

The primary diagnostic feature in olivine is a composite absorption feature at ~1 μm, which consists of three distinct absorption bands. The composite 1 μm band, which is attributed to electronic transitions of $Fe^{2+}$ occupying both the M1 and M2 crystallographic sites (Burns 1970), moves to longer wavelengths as FeO content increases (King and Ridley 1987; Sunshine and Pieters 1998). Pyroxenes have two absorption bands at ~1 μm and ~2 μm that are associated with crystal field transitions in $Fe^{2+}$, which preferentially occupy the M2 site (Clark 1957; Burns 1970) (Fig. 2). Low-calcium pyroxenes, which are conventionally defined as having less than 11 mol% $CaSiO_3$ (wollastonite, or Wo) (Adams 1974), show a well-defined relationship between absorption band positions and composition, as both Band I and Band II positions increase with increasing ferrous iron content (Adams 1974; Burns et al. 1972; Cloutis and Gaffey 1991). There is also a correlation between composition and band positions in high-calcium pyroxene, although the relationship is complicated by the presence of calcium in addition to iron. In spectral type B clinopyroxenes (<Wo50), there is a positive correlation between calcium content and Band I and II position, regardless of iron content (Cloutis and Gaffey 1991). However, there does not appear to be a systematic relationship between Band I or Band II position and Ca and Fe content in spectral type A clinopyroxenes (>Wo50) (Cloutis and Gaffey 1991; Schade et al. 2004). In mixtures of orthopyroxene and clinopyroxene, spectral shape and band positions are dominated by the orthopyroxene component, as clinopyroxene only begins to affect the band positions at abundances >75 wt% (Cloutis and Gaffey 1991).

## 3. NON-COMPOSITIONAL SPECTRAL EFFECTS

Prior to mineralogical interpretation, spectral band parameters of S-type and V-type asteroids must be corrected for non-compositional effects, such as temperature, phase angle, and grain size. Phase angle is defined as the Sun-Target-Observer angle and is typically less than 25° for main belt asteroids but can be much higher for near-Earth asteroids. With increasing or decreasing phase angle the slope of a reflectance spectrum generally becomes redder or bluer respectively, an effect known as phase reddening.

Particle size primarily affects the depth and slope of an absorption feature and overall reflectance, with absorption bands reaching the greatest depth at a grain size bin that provides maximum spectral contrast (deepest absorption bands). Larger particle size typically means deeper bands and blue (more negative) spectral slope and lower overall reflectance. Analysis of particles returned by the Hayabusa spacecraft from near-Earth asteroid (25143) Itokawa have shown that a majority of the 1534 particles have a size range between 3-40 µm, with most of them smaller than 10 µm (Nakamura et al. 2011). This would suggest that spectra of laboratory meteorite samples with grain sizes <100 µm would best represent spectra of asteroid regolith, although the method of sampling likely biased the grain size distribution of the collected material. Temperature differences between asteroid surfaces and room temperature measurements produce changes in absorption band centers, band depths, band widths, and band area ratios. Of these three factors, temperature corrections of band centers have a large impact on mineralogical characterization, whereas phase angle corrections are important for constraining space weathering on small bodies, as they affect spectral slope, albedo, and band depth. Ignoring these effects could lead to erroneous interpretations of asteroid surface mineralogy and space weathering effects (Sanchez et al. 2012).

### 3.1 Temperature-Induced Spectral Effects

The rationale behind applying corrections for temperature-induced spectral effects is motivated by the fact that current laboratory spectral calibrations for the interpretation of asteroid spectra are derived using room temperature mineral/meteorite spectra (~300K). In contrast, the surface temperatures of asteroids can range from 140-440K depending on the object's heliocentric distance. This temperature difference produces changes in absorption band centers, band depths, band widths, and band area ratios (BAR), which in turn can lead to over or underestimation of iron/calcium content in silicates and variations in BAR. Several calibrations for correcting spectral band parameters of specific minerals or meteorite types have been developed in recent years (e.g., Burbine et al. 2009; Reddy et al. 2012; Sanchez et al. 2012).

The first step in temperature correction of asteroid spectral band parameters is an estimation of the sub-solar equilibrium surface temperature, T, (Burbine et al. 2009):

$$T = \left[ \frac{(1-A)L_0}{16\eta\varepsilon\sigma\pi r^2} \right]^{1/4}$$

..............................(Eq. 1)

where $A$ is the bolometric Bond albedo of the asteroid, $L_0$ is the solar luminosity (3.827 x $10^{26}$ W), $\eta$ is the beaming parameter (assumed to be unity) (e.g., Cohen et al. 1998), $\varepsilon$ is the asteroid's infrared emissivity (assumed to be 0.9), $\sigma$ is the Stefan-Boltzman constant (5.67 x $10^{-8}$ J s$^{-1}$ m$^{-2}$ K$^{-4}$), and $r$ is the asteroid's distance from the Sun in meters. If the asteroid albedo $A$ is unknown, a value can be assumed based on its taxonomic classification. The typical variation in the calculated temperature due to changes in the assumed parameters is in the range of +/- 10 K (Burbine et al. 2009).

Burbine et al. (2009) developed equations to correct for temperature-induced spectral effects in V-type asteroids using spectra of bronzite and enstatite measured at various temperatures by Moroz et al. (2000). Reddy et al. (2012) improved those temperature corrections by developing equations based on spectra of Howardite-Eucrite-Diogenite (HED) meteorites over a range of temperatures (Eq. 2-5 in Table 1). These equations help derive a correction factor that is added to the Band I and II centers to obtain values at 300 K. Sanchez et al. (2012) derived similar equations for ordinary chondrites using temperature series from Moroz et al. (2000) and Hinrichs and Lucey (2002). They noted that the Band I center changes due to temperature are negligible and can be essentially ignored (Eq. 6 in Table 1). For monomineralic-olivine or olivine-rich A-type asteroids, Sanchez et al. (2014) derived temperature corrections using data from Hinrichs and Lucey (2002). Similar to previous equations, the Band I center correction is added to the Band I center before analysis (Eq. 8 in Table 1).

Band depth is also affected by changing temperature and causes an increase or decrease in band area ratio. Sanchez et al. (2012) derived equations to correct for temperature effects on BAR of ordinary chondrites. The correction derived using the equation below must be added to the calculated BAR value (Eq. 7 in Table 1). For olivine dominated asteroids, the depth of the 1-μm band changes with temperature. We derived an equation to correct Band I depth (%) for temperature effects using data from Hinrichs and Lucey (2002) (Eq. 9 in Table 1).

### 3.2 Phase Angle-Induced Spectral Effects

Phase angle induced spectral effects can be exhibited as phase reddening, which is characterized by an increase in spectral slope (reddening) and variations in albedo and absorption band depth with changing phase angle (Gradie et al. 1980; Gradie and Veverka, 1986; Clark et al. 2002; Mann et al. 2011; and Shepard and Cloutis 2011) (Fig. 1). This effect, caused by a wavelength dependence of the single scattering albedo, has been observed in both ground based (e.g., Gehrels 1970; Millis et al. 1976; Lumme and Bowell 1981; Nathues et al. 2000, 2010; Reddy et al. 2012) and spacecraft observations (Clark et al. 2002; Bell et al. 2002 and Kitazato et al. 2008) of small bodies. Asteroid observations need to be corrected for phase angle effects because they can influence spectral slope and band depth, which are both non-diagnostic parameters. The equations developed to correct spectral band parameters for phase angle effects normalize them to 0° phase angle.

Reddy et al. (2012) observed Vesta at different phase angles and quantified the effect on spectral slope. Based on these observations they developed an equation to correct the visible spectral slope (0.4-0.7 μm) for phase angle effects:

$$S_c = S - \gamma\alpha \ldots\ldots\ldots\ldots\ldots\ldots\ldots\ldots\ldots\ldots\ldots (Eq.\ 10)$$

where $S_c$ is the phase angle corrected spectral slope, S is the measured spectral slope, $\gamma$ = 0.0198%/μm/deg, and α is the phase angle in degrees. However, data from Dawn predict a similar trend but with less phase reddening (Li et al. 2013) than that observed by Reddy et al. (2012).

Similarly, Reddy et al. (2012) developed an equation to correct the Band Area Ratio for phase reddening based on ground based near-IR spectral observations of Vesta.

$$BAR_c = BAR - \gamma_3 \alpha \ldots\ldots\ldots\ldots\ldots\ldots\ldots\ldots\ldots\ldots\ldots\ldots(Eq.\ 11)$$

where $BAR_c$ is the phase angle corrected BAR, $\gamma_3 = 0.0292/deg$ and $\alpha$ is the phase angle in degrees.

Sanchez et al. (2012) did not find any systematic changes in BAR due to phase angle for ordinary chondrites. They did develop a phase angle correction for correcting band I and II depths of olivine-pyroxene assemblages.

$$BI_{depc} = BI_{dep} - 0.066 \times (\alpha)\ldots\ldots\ldots\ldots\ldots\ldots\ldots(Eq.\ 12)$$
$$BII_{depc} = BII_{dep} - 0.093 \times (\alpha)\ldots\ldots\ldots\ldots\ldots\ldots\ldots(Eq.\ 13)$$

where $BI_{depc}$ and $BII_{depc}$ are the corrected band depths (%) and $\alpha$ is the phase angle. The phase angle range in which these equations are valid are between 2°-70° for Band I and 2°-55° for Band II. While these equations are valid for smaller grain sizes (<45 μm), a plethora of observational data of asteroids and laboratory studies of meteorites have shown that asteroid regolith contains, at a minimum, an optically dominant amount of fine grains.

## 4. FORMULAE FOR DERIVING ASTEROID COMPOSITIONS

### 4.1 Mafic Mineral Compositions

The correlation between mafic silicate compositions (i.e. mol% Fa in olivine and mol% Fs in pyroxene) and band centers is commonly used to determine asteroid compositions from VIS/NIR spectra. Because of the complex nature of the composite olivine band, most work has focused on the correlation between band centers and pyroxene composition. Gaffey et al. (2002) devised a set of equations defining the relationships between band centers and pyroxene compositions, where compositions are expressed as the calcium endmember wollastonite (Wo) and iron endmember ferrosilite (Fs). Though both pyroxene and olivine composition affect the position of Band I, the Gaffey calibrations removed the effect of olivine, thus allowing single pyroxene composition to be measured.

The Gaffey et al. (2002) calibrations, while pioneering, have recently been shown to only work on a limited range of pyroxene compositions and are restricted to mineral assemblages containing a single pyroxene (Gaffey 2007; McCoy et al. 2007). This is a limitation when attempting to use these formulas on meteorite or asteroid mineralogies. For example, no formulas are available for Fs compositions above 50 mol%, which is the range of most Fe-rich eucrites (Burbine et al. 2009). Also, these formulas can greatly overestimate the Fs content in samples containing more than one pyroxene, such as the ordinary chondrites (Gaffey 2007; McCoy et al. 2007). To address the problem of derived pyroxene compositions, Gaffey (2007) developed a procedure to correct for two-pyroxene assemblages. In this procedure, the BAR is used to determine an offset for Band I. The Band II position is corrected by compensating for the presence of augite. This is accomplished by subtracting a predetermined value from the measured Band II position (0.065μm for H chondrites, 0.062 μm for L, and 0.076μm for LL) (Gaffey 2007). The new Band I and Band II centers are then applied to the original Gaffey at al. (2002) equations, resulting in a single low-Ca pyroxene composition for H, L, and LL chondrite analogues. The correct analogue is selected when the derived Fs and Wo values are consistent with measured values.

We tested the Gaffey et al. (2002) equations and subsequent corrections using 38 ordinary chondrites with pyroxene compositions analyzed by Dunn et al. (2010a). As per the guidelines established by the calibration, we selected the equations that satisfied each set of constraints (in terms of Wo and Fs content). In most cases this was equations 2b and 3a from Gaffey et al. (2002). With corrections applied, the difference between measured and derived Wo values ranges from -8.3 mol% to +9.3 mol%, and 63% of derived Wo values are negative. This would suggest the Band I position is being over-corrected. Calculated Fs values are dependent on whether one is testing for H, L, or LL chondrites. Calculated Fs values do not differ substantially between the H and L chondrites, but LL chondrites values are always lower (due to the higher correction factor). The difference between measured Fs values and Fs values calculated using the L chondrite correction range from -7.2 mol% to 9.1 mol%. Often the Fs content is either too low or too high to correctly classify the sample as H, L, or LL. Only 30% of H chondrites, 29% of L chondrites, and 43% of LL chondrites were classified correctly.

Despite the issues associated with using the Gaffey et al. (2002) pyroxene calibrations on ordinary chondrite-like mineralogies, these calibrations are well suited for single pyroxene assemblages, such as the HED meteorites (e.g. Reddy et al. 2010; Hardersen et al. 2011 & 2014). Any calibration for deriving mafic mineral compositions is most successful when designed and implemented for a limited range of mineral assemblages, such as those developed for the Howardites, Eucrites, and Diogenites (HEDs) (Burbine et al. 2007), the ordinary chondrites (Dunn et al. 2010b), and olivine-rich meteorites (Reddy et al. 2011) (Table 2). Each of these sets of calibrations is designed with a specific suite or meteorites and potential asteroid analogues in mind. The Burbine et al. (2007) calibrations (Eq. 11 from Table 2) were developed using a suite of thirteen HED meteorites (basaltic achondrites believed to originate from asteroid (4) Vesta) (McCord et al. 1970). Thus, the Burbine et al. (2007) calibrations are useful for deriving pyroxene chemistries of Vesta-like (V-type) asteroids. These calibrations are preferable to those of Gaffey et al. (2002) because they are simpler, are applicable to the entire range of pyroxene compositions in HEDs ($Fs_{23-56}$; $Wo_{2-14}$), and are more accurate (to within 3 mol% Fs and 1 mol% Wo).

Dunn et al. (2010b) developed formulas for deriving pyroxene and olivine compositions of the ordinary chondrites and asteroids with similar mineralogies (primarily S-types and Q-types). These formulas, one for Fa content in olivine and one for Fs content in pyroxene (Eq. 12 from Table 2), have low error (<2 mol%) and were developed based on the suggestion that Band I center in the ordinary chondrites is controlled almost entirely by the abundance of FeO in olivine and pyroxene and can therefore serve as a proxy for bulk FeO content. This also suggests that there is no significant correlation between Band I center and CaO content (mol % Wo) in meteorites and asteroids that consist of both olivine and pyroxene. However, we would expect CaO content to have a much more significant effect on Band positions in pyroxene-dominated asteroids, such as the vestoids.

Recently, Reddy et al. (2011) added another formula to derive forsterite (Fo) content of olivine in meteorites and asteroids composed almost entirely of olivine (Eq. 13 from Table 2). The relationship between Band position and olivine composition was first examined by King and Ridley (1987). Reddy et al. (2011) improved this calibration by adding more Mg-rich olivines (Fo85-93) to the sample suite. The root mean squared error between measured values and laboratory derived Fo-content is 5 mol%. This equation is designed to be used with A-type asteroids, which have a weak or nearly absent 2-µm feature consistent with a monomineralic olivine composition.

One must use caution when applying any of these calibrations to asteroids, as inaccurate derived mineralogies can lead to misinterpretation of asteroid geologic histories. Such is the case of Itokawa, the target of the Hayabusa mission (see Section 6). Abell et al. (2007) suggested that Itokawa has experienced partial melting, based on its high-Fe pyroxene content, while others argued it was compositionally similar to the ordinary chondrites (e.g., Binzel et al. 2001; Abe et al. 2006; Okada et al. 2006). Itokawa's primitive origin (as an LL chondrite) was confirmed by the Hayabusa sample return mission (Nakamura et al. 2011). However, this extremely rare case of obtaining the "ground truth" for specific asteroids is only possible via sample return or terrestrial impact and recovery of an observed asteroid, such as 2008 $TC_3$.

Figure 3 shows a decision-making flowchart for interpretation of asteroid spectra based on mineralogical calibrations that are available. In asteroids that show only a 1-µm absorption feature, iron abundance in olivine can be constrained using its Band I center position and equations from Reddy et al. (2011). For asteroids with 1- and 2-µm bands, similar calibrations exists depending on where the asteroid spectral band parameters plot on the Band I center vs. Band Area Ratio (BAR) plot (Gaffey et al. 1993). This Band I center vs. BAR plot is broken into S-type subclasses with different zones marking different compositional types. Asteroids in the S(IV) region can be mineralogically characterized using Dunn et al. (2010) calibrations. Those that fall in the basaltic achondrite zone can be interpreted using calibrations from Gaffey et al. (2002) and Burbine et al. (2009). Meteorite analogs for these three asteroid types (olivine-rich A-type asteroids; S/Q-type asteroids, V-type asteroids) constitute 91% of all the meteorites in our terrestrial collection (DeMeo et al. in this book).

**4.2 Mafic Mineral Abundances**

In spectra containing both olivine and pyroxene absorptions, the combined absorption features near 1 µm (Band I) and near 2 µm (Band II) are also sensitive to the relative proportions of olivine and pyroxene. The ratio of the areas of these two bands (Band II/Band I) is commonly used to estimate olivine and pyroxene abundances in meteorites and asteroids (Cloutis et al. 1986). The linear relationship between this band area ratio (BAR) and the ratio of pyroxene to olivine + pyroxene (px/(ol+px)) was first recognized by Cloutis et al. (1986). However, the mixtures Cloutis et al. (1986) used to derive their equation did not contain clinopyroxene. As a result, their calibrations were not well-suited for determining the BARs of mixtures containing more than one pyroxene (i.e. ordinary chondrites).

The Cloutis et al. (1986) calibration was initially designed to derive BARs from mixtures of known mineral proportions. However, this relationship between mafic mineral abundances and BAR is also an important tool for determining asteroid mineralogies from VIS/NIR spectra. Gastineau-Lyons et al. (2002) used this relationship to derive mineral abundances from measured BARs of asteroid spectra. However, because the original Cloutis et al. (1986) regression was based on simple mixtures of olivine and orthopyroxene, the presence of more than one pyroxene (or other additional phases) complicated spectral interpretations of asteroids made using this calibration (Gaffey et al. 1993; Sunshine et al. 2004).

In an attempt to develop BAR calibrations that were more useful in asteroid spectroscopy, Burbine et al. (2003) used normative abundances of the ordinary chondrites (abundances calculated from measured bulk chemistry) to derive a calibration for the S-type asteroids. The most recent revision of the calibrations for S-type asteroid mineralogies was produced by Dunn et al. (2010b). Dunn et al. (2010b) used measured modal abundances of ~50 ordinary chondrite powders to derive their calibrations. Modes were determined using a position sensitive X-ray diffraction technique (Dunn et al. 2010c), which allows for measurement of phases present at abundances as low as 1 wt% (Cressey et al. 1996). Using this technique, Dunn et al. (2010b) were able to account for pigeonite, the third pyroxene present in the ordinary chondrites. Because this regression was derived using measured abundances (as opposed to abundances calculated using CIPW norms), it provides a more accurate measure of olivine and pyroxene abundances from VIS/NIR spectra.

Gaffey et al. (2002) suggested that pyroxene abundances could be constrained even further by considering the effect of high-Ca pyroxene on Band II position, which they suggested is a linear function of the relative abundance of the two pyroxenes. This assertion requires that the cpx/(opx+cpx) ratio is well-correlated with Band II center in the range of values found in ordinary chondrites. Using XRD-measured pyroxene abundances and Band II centers, Dunn et al. (2010b) derived an equation for establishing the ratio of high-Ca pyroxene to total pyroxene. However, this equation yielded a very low $R^2$ value of 0.15. The remarkably low $R^2$ value indicates that there is no significant correlation between Band II position and relative pyroxene abundance in the ordinary chondrites.

Though the Dunn et al. (2010b) regression formula is well suited for asteroids with ordinary chondrite-like compositions (e.g. some S-types and Q-types), it is not as equally well-suited for asteroids with other mineralogies. For example, the original calibration of Cloutis et al. (1986) is still better suited for asteroids that are compositionally similar to primitive achondrites (lodranites and acapulcoites), which are primarily mixtures of low-Ca pyroxene and olivine (e.g., McCoy et al. 2000). The Dunn et al. (2010b) calibrations would not work well for mineralogies that are dominated by only one mafic phase (i.e. either olivine or pyroxene, but not both). Recently, Reddy et al. (2011) developed a calibration for olivine-rich asteroids using the R chondrites (Eq. 14 from Table 2), which contain higher abundances of olivine than the ordinary chondrites (65-78%) (Bischoff et al. 2011). Initial tests of this calibration yield abundances in strong agreement with previous characterizations. However, it is important to note that this calibration was developed using a limited sample selection (only seven R chondrites). Also, though R chondrites provide a good spectral match with olivine-dominated asteroids, their compositions may differ (Reddy et al. 2011). Consequently, any interpretations made using the Reddy et al. (2011) formula should be taken with caution. Additional laboratory measurements will be necessary to improve the robustness of this calibration.

# 5. SPECTRAL MODELING

## 5.1 Modified Gaussian Model

Composite mineral absorption bands can be de-convolved into discrete mathematical functions. The modified Gaussian Model or MGM (Sunshine et al., 1990) constructs a model spectrum from one or more modified Gaussian functions to represent absorption features from individual species (Fig. 4). These modified Gaussians are physically motivated by Crystal Field Theory and represent the relationship between the natural log of reflectance and energy (Sunshine et al., 1990). In comparison to ordinary Gaussians these modified functions include the addition of a power law index of $n=-1$ to their energy dependence. Each MGM band is associated with three degrees of freedom (center, width, depth). The full model includes a continuum, which is typically linear in energy space and adds two free parameters for slope and offset. Initial values for all of these parameters are specified at the onset of modeling and are iteratively adjusted until a best fit is determined by minimizing the RMS residuals of the difference between reflectance data and the model spectrum.

The MGM was originally employed to analyze laboratory spectra of terrestrial and meteorite samples (Sunshine et al., 1990; Sunshine and Pieters, 1993) and has since been extensively applied to mafic silicate systems including the Moon (e.g. Isaacson et al., 2011), Mars (e.g. Clenet et al., 2013), and a variety of asteroid types. In its most basic implementation, the MGM can be used as a simple band analysis tool to measure band parameters, analogous to those discussed in Section 3. Using the MGM to derive band parameters is beneficial because it can serve as a consistent analysis tool that does not vary from one implementation to the next. This is not necessarily the case with other band analysis techniques that may employ different methodologies (e.g. Cloutis et al., 1986; Moskovitz et al., 2010; Dunn et al., 2013). Standardizing a tool like the MGM for measuring band parameters can facilitate comparisons across investigations, but does require that continua be modeled in a consistent manner.

The MGM has been employed as a band analysis tool in several studies of near-Earth asteroids. Rivkin et al. (2004) showed that the MGM could reliably recover Band 1 areas for S-type NEAs in cases where only near-IR spectra were available and the 1 μm band was not fully resolved. These authors defined the MGM-derived "RBA" (ratio of band areas) parameter as an analog to BAR for spectra where visible data were not available. Though not extensively tested, the RBA reasonably reproduces compositions for objects with well-constrained mineralogy like Itokawa and Eros. Vernazza et al. (2008) and Thomas and Binzel (2010) extended the use of the MGM as a band analysis tool to link H, L and LL ordinary chondrites to their asteroidal counterparts in near-Earth space. The MGM has also been used to correlate the band parameters of V-type asteroids with subgroups of the HED meteorites (Mayne et al., 2011).

In its full implementation the MGM is used to model specific mineral absorption bands. This can involve diagnosing the origin of unknown features (e.g. Hiroi et al., 1996) or starting with *a priori* assumptions about composition to generate a mineralogical model (e.g. Mayne et al., 2011). For example, the relative strengths of pyroxene and olivine bands in the vicinity of 1 μm led Binzel et al. (2001) to correctly predict an LL chondrite-like composition for (25143) Itokawa (Fig. 4). Analyses of other S-complex asteroids suggest that the MGM can constrain the relative abundance of high-Ca to low-Ca pyroxene. For example, Sunshine et al. (2004) found high-Ca pyroxene abundances >40% for the large Main Belt asteroids (17) Thetis, (847) Agnia, and (808) Merxia, thus implying that these objects experienced high degrees of partial melting consistent with a history of igneous differentiation.

The three overlapping 1 μm olivine bands, as seen in the spectra of A-type asteroids, are particularly well suited to MGM analysis, despite their sensitivity to space weathering (e.g. Brunetto et al., 2007) and temperature (e.g. Lucey et al., 1998). Using the MGM, Sunshine et al. (2007) showed that A-type asteroids segregate into two groups based on olivine composition. Those with magnesian (forsteritic) compositions suggest an igneous history involving the differentiation of ordinary chondrite precursors, whereas those with ferroan (fayalitic) compositions are more consistent with an origin tied to oxidized R-chondrite precursors. K-type asteroids in the Eos family have also been associated with an olivine-dominated composition (Mothe-Diniz et al., 2008).

Due to their prominent 1- and 2-μm absorption features, pyroxene-dominated V-type asteroids have been extensively modeled with the MGM. Traditional band analyses of these asteroids (Section 3) have generally focused on deriving bulk pyroxene composition (e.g. Wo, Fs, En #'s), whereas MGM analyses are typically used to measure the abundance ratio of high-Ca to low-Ca pyroxene. These different approaches can provide complementary insight when performed in tandem (e.g. *Lim et al.,* 2011). MGM analyses of V-type asteroids suggest that very roughly 25-40% of their total pyroxene is in the high-Ca phase (Duffard et al., 2006; Mayne et al., 2011), consistent with the pyroxene composition of the HED meteorites (Mittlefehldt et al., 1998) and therefore supporting the generally accepted link between these asteroids and meteorites.

Like any compositional analysis tool the MGM has limitations. Perhaps its greatest weaknesses are the dependence on multiple free parameters in even relatively simple mineralogical models and the sensitivity of those parameters to prescribed initial conditions. Without careful application these issues can produce non-diagnostic or degenerate fits.

### 5.2 Radiative Transfer Models

Models of radiative transfer on atmosphereless bodies generally involve computing reflectivity (i.e. scattering and phase corrected single-scattering albedo) based on optical constants and particle sizes of specific mineral species. Reflectivity is computed as a function of assumed viewing geometry and is rooted in approximations of geometric optics. Radiative transfer models can provide insight into particle sizes, mineral compositions, and mineral abundances.

Various approaches have been developed to treat the presumably chaotic structure and mixture of mineral grains in planetary regoliths. The approximation of an areal or linear mixture involves computing the reflectance from mineral phases separately and then linearly combining those to represent the total reflectance. This "checkerboard" approach assumes the constituent minerals are optically separated so that multiple scattering does not occur between minerals. Intimate, non-linear, or "salt-and-pepper" mixtures involve computing an average single-scattering albedo of the mineral constituents as input to the reflectivity calculation. The sizes of particles relative to the incident wavelength of light determine how the single-scattering albedo is calculated in such homogeneous mixtures (Shkuratov et al., 1999, Poulet et al., 2002).

Radiative transfer models are some of the most widely used tools for compositional analyses of planetary bodies. The theory outlined by Hapke (1981) and since improved upon in a number of subsequent publications (*Hapke* 1984; 1986; 2001; 2002; 2008) was one of the first to derive the single scattering albedo for a grain with prescribed size and optical properties. Shkuratov et al. (1999) developed a similar but alternative model that employs a distinct treatment of particle scattering. A detailed review and comparison of these two implementations is presented in Poulet et al. (2002). More recently Lawrence and Lucey (2007) expanded upon the Hapke formalism by incorporating the spectral effects of coarse-grained Fe-Ni metal to facilitate modeling of the metal-rich assemblages frequently encountered in small body studies, thus enabling studies of the effects of space weathering on mafic silicate bodies like the Moon, basaltic achondrites, and ordinary chondrites. Finally, Lucey and Riner (2011) introduced a new treatment of large, opaque iron agglutinates as a way to model the optical effects of weathering via impact and shock alteration processes.

Such radiative transfer models have been used for a variety of asteroid applications. Based on spectro-photometry from the Near Earth Asteroid Rendezvous (NEAR) mission, Clark et al. (2001; 2002) modeled near-IR spectra of the S-type NEA (433) Eros and were able to disentangle viewing geometry, grain size, composition, and surface weathering effects as causes of color and albedo variability across the asteroid's surface. Clark et al. (2004) applied the same model to E-type asteroids and found that this single taxonomic class could actually be broken into three compositional groups: objects like (44) Nysa with enstatite-like compositions that include a low-iron orthopyroxene component, objects like (64) Angelina composed of forsterite olivine and oldhamite, and objects like (434) Hungaria, which may be linked to the enstatite-rich aubrite meteorites.

Several variants of radiative transfer models have been used to address issues related to space weathering. Employing a variant of the Shkuratov model, Vernazza et al. (2008) showed that the majority of S-type NEAs, in spite of their weathered appearance, have compositions most similar to LL ordinary chondrites (Fig. 5). This is surprising in light of the fact that LL chondrites are rare amongst ordinary chondrite meteorite falls and may be a result of selection bias influencing the delivery of meteorites from the Main Belt to Earth. This same model was employed to argue that the timescale of space weathering is a rapid $\sim 10^6$ years (Brunetto et al., 2006; Vernazza et al., 2009a); however, this result remains somewhat controversial (Willman and Jedicke, 2011). As such, it is currently unclear how effective these models are in connecting the timescales of space weathering to observed spectral changes.

This particular implementation of the Shkuratov model (Vernazza et al. 2008) does not attempt to fit data beyond ~2 µm due to ineffectiveness of the applied space-weathering model (Fig. 5). However, this implementation has been used exclusively to derive olivine to pyroxene abundance ratios, information which is entirely contained in the 0.6-1.6 µm spectral range and therefore is not dependent on the properties of the 2 µm band. As such, future work to derive silicate abundances could benefit from exclusive focus on spectral information contained within the 1 µm region.

Primitive compositions have also been modeled with radiative transfer techniques. The spectra of primitive asteroid types (e.g. C- and D-types) are generally red sloped and largely devoid of prominent absorption features, particularly at visible and near-IR wavelengths. Thus, compositional models of these objects can be less specifically diagnostic, instead providing upper limits on mineral abundances. For example, Cruikshank et al. (2001) argued that the surface of the Trojan asteroid (624) Hektor contained <3 wt% water ice and <40% hydrated silicates. Their models further suggested that an unknown low albedo, spectrally neutral material, potentially analogous to elemental carbon, is responsible for the red color and low albedo of D-type asteroids. Emery and Brown (2004) expanded upon this work by modeling the spectra of 17 Trojans. Their results are consistent with *Cruikshank et al.* (2001), suggesting that the Trojan's red spectral slopes are unlikely due to organics and that the Trojan's overall spectral properties may be attributed to a combination of anhydrous silicates and carbonaceous material. Based on data at thermal-IR wavelengths, Emery et al. (2006) modeled the emissivity spectra of several Trojans to argue for compositions due to fine grained silicates embedded within an optically transparent matrix. Yang and Jewitt (2011) largely corroborated this result and suggested that Trojan silicates should be iron-poor to account for spectral data at both near-IR and thermal-IR wavelengths. Finally, Yang et al. (2013) modeled Trojan spectra to suggest a composition with small quantities of fine grained silicates (1-5 wt%) and a low albedo component (2-10 wt%) embedded within a matrix of salt deposits. In spite of these advances, full compositional characterization of Trojan asteroids remains an unsolved problem.

Perhaps the best examples of featureless spectra that evade compositional characterization are those belonging to the asteroid (21) Lutetia. Even though Lutetia was the target of a spacecraft flyby (Barrucci et al., this volume), radiative transfer models suggest its composition consists of weathered goethite (Rivkin et al., 2011), a hydrated iron-bearing mineral, whereas spectral matching to laboratory samples suggests a very different composition, more analogous to enstatite chondrites (Vernazza et al., 2009b).

Spectral data at longer wavelengths (> 3 µm) can better diagnose specific compositions on primitive asteroids. For example, radiative transfer models have provided specific insight into the presence of hydrated minerals on primitive asteroids (e.g. *Rivkin et al.,* 2006; *Rivkin and Emery,* 2010; *Campins et al.,* 2010). Compositional analysis at mid-IR wavelengths remains a major frontier of asteroid science (See Rivkin et al. chapter in this book).

### 6. COMPOSITIONAL CONSTRAINTS FROM MID-IR OBSERVATIONS

In the mid-infrared wavelength region spectral bands are chiefly due to vibrational and lattice modes. The resulting reflectance and emission spectra in these wavelengths are strongly dependent on surface parameters such as grain size (e.g., Lyon 1964, Hunt & Logan 1972, Salisbury et al. 1991a, Le Bras & Erard 2003) and porosity (Aronson et al. 1966, Logan et al. 1973, Salisbury et al. 1991b), as well as environmental conditions such as ambient pressure (Logan et al. 1973, Salisbury & Walter 1989)

## 6.1 Diagnostic Spectral Features in the Mid-IR

Laboratory studies of powdered meteorites and silicate minerals display a variety of diagnostic spectral signatures: Christiansen features, Restrahlen bands, transparency features, and absorption bands. The Christiansen feature is a reflectance minimum (emissivity maximum) located ~ 7.5-10 μm in most silicates. In this wavelength region the real part of the refractive index changes rapidly (dropping through unity) and therefore may approach the refractive index of the medium surrounding the mineral grains resulting in minimal scattering. The location where the real refractive index equals unity is the Christiansen Frequency (e.g., Logan et al. 1973). For silicates, the Christiansen Frequency occurs at a shorter wavelength than the Restrahlen band where scattering is minimal and the absorption coefficient is small. The resulting reflectance minimum (the Christiansen feature) is often located at longer wavelengths then the Christiansen Frequency. Hapke (1996) interpreted this feature as occurring due to a transition between the surface and volume scattering regimes. Past work (Conel 1969, Logan et al. 1973, Salisbury & Walter 1989) has noted that the wavelength of the Christiansen feature is an indicator of mineral composition. The wavelength of the Christiansen feature for individual minerals depends on the polymerization of the mineral with the feature occurring at shorter wavelengths for feldspars and longer wavelengths for mafic minerals such as olivine and pyroxene (Logan et al. 1973). For a given asteroid or meteorite, the Christiansen feature is the combination of the characteristic Christiansen features of all the constituent minerals. Often times the principal Christiansen feature corresponds to the strongest molecular vibration band. However, in the case of some meteorites, the individual Christiansen features are far enough apart that numerous local reflectance minima are seen.

Restrahlen bands are the result of strong fundamental molecular vibrations. For silicates they appear at ~8.5-12 microns and are the product of Si-O asymmetric stretching (e.g., Lyon 1964, Hunt 1982, Salisbury et al. 1991). These bands appear as peaks in reflectance and troughs in transmittance (Vincent and Hunt 1968) and are centered on the short wavelength side of the maximum of the imaginary index. In this wavelength region the absorption coefficient and specular reflectance are high (Salisbury et al. 1987) causing most incident radiation not to enter the sample and instead be reflected on the first surface. Lyon (1964) showed that restrahlen bands are difficult to detect in the spectra of particulate minerals because of multiple occurring reflections, which increase the proportion of photons that enter the medium and are absorbed. Therefore, the height of restrahlen bands decreases with decreasing grain size (Salisbury et al. 1987, Le Bras et al. 2003). A second group of less intense restrahlen bands occurs between ~16.5 and 25 microns and are associated with Si-O-Si bending. Restrahlen features of silicates shift to longer wavelengths when heavier cations are in the lattice structure since they slow the frequency of the molecular vibrations. Meteorite spectra (Salisbury et al. 1991) show that the restrahlen region is dominated by olivine and pyroxene. These spectra from Salisbury et al. (1991) also demonstrate the slight shift to longer wavelengths with increasing iron content that can be seen in the ordinary chondrite's progression from more forsteritic olivine (H chondrites) to more fayalitic olivine (LL chondrites).

Between the two restrahlen bands, the spectra of particulate silicates are dominated by a broad reflectance maximum (emissivity minimum) known as the transparency feature. The absorption coefficient is low enough in this region that silicate grains are optically thin and volume scattering dominates the scattering process. The feature is dependent on the presence of small particles (< 75 microns) and changing optical constants (Salisbury et al. 1987b, Salisbury & Walter 1989, Mustard & Hays 1997). Laboratory spectra of silicate minerals demonstrate that the wavelengths of the transparency peaks and Christiansen features are diagnostic of composition and are correlated with each other (Logan et al. 1973, Salisbury et al. 1987b, Salisbury & Walter 1989, Hapke 2012).

Absorption features in the mid-IR are present at various wavelengths. For primitive carbonaceous chondrites, the most prominent absorption feature is the O-H stretching vibration feature near 3.0 microns (Salisbury & Hunt 1974, Miyamoto 1988). This absorption feature tends to increase in spectral contrast as particle size is reduced below 75 microns since the continuum reflectance rises with increasing scattering (Salisbury & Walter 1989, Salisbury et al. 1991).

Crystalline olivines have strong diagnostic absorption features between 9 and 12 microns (due to Si-O stretches) as well as several other minor bands at longer wavelengths. All olivine bands shift to shorter wavelengths with decreasing Fe/(Fe+Mg) content (Koike et al. 2003) and decreasing temperature (Kioke et al. 2006). Crystalline pyroxenes have absorption features at similar but slightly shorter wavelengths than the olivines. Most of these features shift to shorter wavelengths with decreasing Fe/(Fe+Mg) content, except the 10.5 and 11.5 micron bands, which shift to longer wavelengths (Chihara et al. 2002). Past spectral studies of crystalline olivine and pyroxene under a variety of laboratory conditions include Mukai & Koike (1990), Koike et al. (1993) Jager et al. (1998), Henning and Mutschke (1997), Pitman et al. (2010), and Lane et al. (2011). Amorphous olivines and pyroxenes have broad absorption features near 10 microns with a weaker band near 20 microns. Laboratory investigations of these amorphous silicates include Day (1981), Koike & Hasegawa (1987), Scott & Duley (1996), and Brucato et al. (1999).

Mid-infrared (or thermal) reflectance spectra (5-25 µm) of crystalline olivines and pyroxenes with various compositions and grain sizes are available at the ASTER spectral library (Baldridge et al. 2009; http://speclib.jpl.nasa.gov). Additional data, including optical constants and references, are available at JPDOC (Henning et al. 1999; http://www.astro.uni-jena.de/Laboratory/Database/jpdoc/f-dbase.html).

### 6.2 Astronomical Observations in the Mid-IR

One key difference between mid-infrared spectral laboratory data and mid-infrared astronomical observations of asteroids is surface pressure. Logan et al. (1973) first demonstrated that decreasing the atmospheric pressure for a particulate sample results in an increase in spectral contrast, a shift to shorter wavelengths of the

Christiansen feature, and potential loss of the transparency feature. These spectral changes are due to the fact that the space environment introduces thermal gradients in a particulate sample that alters the spectral features. At lower pressures a sharp thermal gradient occurs, as less or almost no interstitial gas is present to convectively transport heat. In a vacuum, the sample is heated by the absorption of radiation in visible wavelengths to a depth that is dependent on the absorption coefficient in visible wavelengths, and the sample emits to a cooler background in infrared wavelengths over a depth dependent on the absorption coefficient in infrared wavelengths. For silicates, the absorption coefficient for visible wavelengths is much smaller than the absorption coefficient in near-infrared wavelengths. This results in an emitting layer of particles that is shallower than the absorbing layer. Different wavelength regions in the spectrum are effectively sampling the two different temperatures, which increases the emissivity contrast.

In addition to surface pressure, grain size effects can also hinder mid-infrared astronomical observations of asteroids. Various authors (e.g., Lyon 1964, Salisbury et al. 1987, Le Bras et al. 2003) have demonstrated that the prominent restrahlen features are difficult to detect for particulate surfaces and that the strength of the restrahlen features varies inversely proportionally to particle size. Grain size effects can also impact attempts to apply spectral deconvolution techniques to mid-infrared data. Past work has demonstrated that mid-infrared spectra represent linear combinations of the abundances of the individual components (Lyon 1964, Thomson & Salisbury 1993, Ramsey & Christensen 1998, Hamilton & Christensen 2000). The effectiveness of linear spectral deconvolution techniques for asteroid surfaces is dependent on the particle size of the regolith. Ramsey & Christensen (1998) found that linear deconvolution fails for grain sizes below 10 μm where non-linear volume scattering dominates.

Finally, another important factor to consider when using astronomical observations of asteroids to determine silicate mineralogy is the presence of significant Si-O absorption features in the mid-infrared spectra of observed standard stars. Cohen et al. (1992) identified the Si-O fundamental absorption in stellar spectra. Prior to that, early work to investigate asteroids using mid-infrared spectroscopy was hindered by insufficient understanding of the nature of the standard stars (e.g., Gillet and Merrill 1975, Green et al. 1985). The Kuiper Airborne Observatory (KAO) and the ESA Infrared Space Observatory (ISO) observed several large asteroids in the 1990s. KAO spectra (5-14 microns) of (1) Ceres showed three potential emission features, but no mineralogical match was identified (Cohen et al. 1998). The ISO spectra (6-12 microns) showed evidence of silicates on the surface (Dotto et al. 2000) and made connections between (10) Hygiea and (511) Davida and the CO and CM carbonaceous chondrites (Barucci et al. 2002, Dotto et al. 2002).

## 6.3 Recent Advances in Mid-IR Asteroid Spectroscopy

Since Asteroids III, there have been several notable works connecting mid-infrared spectroscopy of asteroids to meteorites and silicate mineralogies. Much of the work in this area has focused on the 3-μm region where hydrated silicates have a distinct absorption feature. This spectral feature was discussed in great detail in Asteroids III (Rivkin et al. 2002). Recent work by Takir & Emery (2012) classified 3 μm asteroid spectra into four distinct groups, demonstrating that there was thermal stratification of outer Main Belt asteroids early in the history of the Solar System. The asteroids that had experienced aqueous alteration (the "sharp" group) are located in the 2.5<a<3.3 AU region, while those that remained unaltered (the "rounded" group) are in the 3.3<a< 4.0 AU region.

Lim et al. (2005) performed a survey of 29 asteroids using 8-13 μm spectra. Their analysis identified emissivity features for (1) Ceres, which were similar to those seen by Cohen et al. (1998) with the KAO and Dotto et al. (2000) with ISO. No definitive mineralogy was established. They also noted a marginal detection of features for (4) Vesta, which matched features found in HED meteorites (Salisbury et al. 1991). Work by Lim et al. (2011) attempted to establish a formalism for connecting features in the emissivity spectra of the V-type asteroid 956 Elisa to those seen in the spectra of HED meteorites.

Emery et al. (2006) presented mid-infrared (5.2 - 38 μm) spectra of three Trojan asteroids: (624) Hektor, (911) Agamemnon, and (1172) Aneas. Using spectra of fine-grained silicates from the ASTER spectral library they concluded that the spectra are characteristic of either small silicate grains imbedded in a relatively transparent matrix or small silicate grains in a very under-dense (fairy castle) surface layer. Mid-infrared (8-33 μm) spectra of (617) Patroclus by Mueller et al. (2010) showed similar emissivity features, suggesting that the object has similar mineralogy to the other observed Trojans.

The work of Vernazza et al. (2010) showed spectral variability between mid-infrared spectra of eight S-type asteroids, which previous analyses using visible and near-infrared spectroscopic techniques had determined were of similar compositions. The analysis concluded that the variability was due to differences in the regolith particle size and space weathering effects. The authors indicated that the asteroids are covered with a very fine-grained regolith (<5 microns) that conceals some spectral features.

Vernazza et al. (2012) noted that while emission features on asteroids in the mid-infrared spectral region were common, the asteroid spectra do not always match the laboratory spectra of minerals and meteorites. The authors demonstrated that by suspending the meteorite and mineral powders in potassium bromide powder, which is transparent in the infrared, and were able to recreate the spectra of Main Belt asteroids with emissivity spectra. The work suggests that these asteroids have extremely porous regolith layers.

Irradiation experiments of the Tagish Lake meteorite demonstrated a small shift to longer wavelengths of the 10 μm silicate emission feature that is likely caused by the amorphization of the silicates (Vernazza et al. 2013). They used a combination of albedo with visible, near-infrared, and mid-infrared spectra to determine that Tagish Lake-like objects are a very small percentage of the main belt population and that (368) Haidea is the best match to the meteorite's spectrum and albedo.

## 7. CONSTRAINTS FROM SPACECRAFT MISSIONS

Starting with the Galileo flyby of asteroids (951) Gaspra and (243) Ida, spacecraft investigations of asteroids have provided validation for ground-based spectroscopic studies. More recently, the Japanese Hayabusa and NASA's Dawn missions have completed detailed mineralogical studies of asteroids (25143) Itokawa and (4) Vesta, respectively, from orbit. The Hayabusa mission has also brought back microscopic samples of Itokawa that have helped immensely in the confirmation of precursor telescopic observations. Initially the surface composition of Eros was ambiguous, with derived compositions ranging from H-L-LL chondrites to primitive achondrites (McFadden et al. 2001; Izenberg et al. 2003; McCoy et al. 2002). Further analysis of spectral observations of Eros by the Near-Infrared Spectrometer (NIS) and Multispectral Imager (MIS) on NEAR suggested a surface composition consistent with a space-weathered L6 ordinary chondrite (Nittler et al. 2001; McCoy et al. 2002; Izenberg et al. 2003). McFadden et al. (2005) reported evidence for partial melting based on Modified-Gaussian model (MGM) analysis of Eros reflectance spectra. This debate between partial melt vs. space weathered L6 chondrite was further fueled by the results from the NEAR X-Ray Spectrometer (XRS) experiment, which discovered strong sulfur depletion. Nittler et al. (2001) argued that the best explanation for the depletion is space weathering on an ordinary chondrite type object.

Similar to Eros, the surface composition of near-Earth asteroid (25143) Itokawa was the subject of much debate prior to the arrival of the Hayabusa spacecraft at the target object and subsequent return of the asteroid's samples to Earth. Binzel et al. (2001) interpreted the surface composition of Itokawa to be a space weathered LL chondrite with 0.05 % nanophase iron on the surface. In contrast, Abell et al. (2007) argued that the mean pyroxene chemistry of Itokawa (~$Fs_{43±5}Wo_{14±5}$), estimated from its spectrum using calibrations from Gaffey et al. (2002), indicated a surface that experienced partial melting. Analysis of samples returned by the Hayabusa spacecraft (Nakamura et al. 2011) has shown that Itokawa is indeed an LL chondrite as suggested by Binzel et al. (2001). Mineralogical analysis of the returned samples and the availability of newer spectral calibrations from Dunn et al. (2010) provide us with an opportunity to validate ground-based spectral interpretive tools.

We used the spectral band parameters of Itokawa from the NASA Infrared Telescope Facility (Binzel et al. 2001; Abell et al. 2007) to calculate its olivine and pyroxene ratios and chemistries and to compare them with laboratory-measured values from samples returned by the Hayabusa spacecraft. Figure 6 (reproduced from Nakamura et al., 2011) shows olivine iron abundance (fayalite) on the Y-axis and pyroxene iron abundance (ferrosilite) on the X-axis from laboratory measurements of ordinary chondrites (H, L and LL), Itokawa samples returned by Hayabusa spacecraft (filled circle), and those derived from the ground based telescopic spectrum (open circle) using the Dunn et al. (2010) calibrations. The difference between laboratory measured values of Itokawa samples and those from ground-based spectral data is less than 1 mol. %, confirming Itokawa as an LL chondrite and attesting to the validity of the Dunn et al. (2010) calibrations. Similarly, we used spectral band parameters for Eros from the NIS instrument on NEAR spacecraft (Izenberg et al. 2003) and ground-based spectral data (Murchie and Pieters 1996) to calculate its olivine and pyroxene chemistries to a) check for consistency between the two data sets, and b) validate the L chondrite interpretation by Izenberg et al. (2001). Figure 6 also shows fayalite and ferrosilite values for Eros calculated from NEAR and ground-based spectral data. These results validate Izenberg et al. (2001) interpretation that Eros surface composition is similar to an L chondrite rather than a partial melt, as proposed by McFadden et al. (2005).

NASA's Dawn spacecraft completed a year-long mission to characterize asteroid (4) Vesta (Russell et al. 2012), mapping the surface with a multi-color Framing Camera (FC) (Reddy et al. 2012, Jaumann et al. 2012), visible-near-IR spectrometer (VIR) (De Sanctis et al. 2012), and Gamma-ray and Neutron detector (Prettyman et al. 2012). Prior to Dawn's arrival, Vesta had been the most studied asteroid in the asteroid belt (e.g., Gaffey, 1997; Reddy et al. 2010; Thomas et al. 1997a; Binzel et al. 1997; Li et al. 2010) with nearly 100 years of spectral observations dating back to the early part of last century (Bobrovnikoff, 1929). Many ground-based and HST observations of the Vestan surface were confirmed by the Dawn mission, such as the hemispherical albedo dichotomy between the eastern and western hemisphere (Reddy et al. 2013), which was proposed based on monomodal photometric lightcurve observations of Vesta (Li et al. 2010). Like many observations, this was confirmed by comparing ground-based and Hubble Space Telescope (HST) albedo and compositional maps of Vesta (Gaffey 1997; Binzel et al. 1997; Li et al. 2010) to multi-color camera data from the Dawn mission (Reddy et al. 2010). Figure 7 shows albedo features identified first in HST images (Binzel et al. 1997; Li et al. 2010) and then by the Dawn spacecraft. Features are annotated to show positive identification of geological features on both data sets.

While ground-based and HST observations of Vesta matched with those made by Dawn, the interpretation of certain features was enhanced due to better spatial resolution of the spacecraft data. Of particular interest to the ground-based spectroscopy community was the 'olivine-rich' unit named "Leslie feature" in the Gaffey (1997) IRTF map of Vesta. Rotationally resolved spectra of Vesta showed a drop in Band Area Ratio (BAR) when this feature rotated into view. Le Corre et al. (2013) identified this as the Oppia crater and its surrounding ejecta blanket but suggested that this drop in BAR was due to the presence of impact melt in the ejecta blanket rather than olivine. The rotational compositional variation between the northern and southern hemispheres, where the South Pole Rheasilvia basin excavated more diogenetic material (Reddy et al. 2010), was also confirmed by Dawn FC and VIR data (Reddy et al. 2012; De Sanctis et al. 2012). Dawn observations also confirmed the Hasegawa et al. (2003) hypothesis that the Vestan surface was contaminated by carbonaceous chondrite impactors (Reddy et al. 2012b; McCord et al. 2012; De Sanctis et al. 2012b; Prettyman et al. 2012). This hypothesis was initially based on the ground-based identification of an OH absorption band at 2.8 μm. Results from the Dawn mission have provided the ground-based spectroscopy community an opportunity to verify nearly a century of observations of Vesta.

## 8. ASTEROIDS WITH DERIVED MINERALOGIES

The challenges to link a meteorite with a particular parent body or source body are daunting. A parent body is the original body that the meteorite formed on, while the source body is a larger fragment of the original parent body that is the direct source of the meteorite. Lunar meteorites were easily identified (e.g., Marvin 1983) due their similarity in mineralogy and chemistry to returned Apollo samples. Martian meteorites were identified (e.g., Bogard and Johnson 1983), in part, by the similarity of gas abundances found in meteoritic glasses and the composition of the Martian atmosphere analyzed by the Viking spacecraft. The only instance that a link between a meteorite and its asteroid source has been made with complete certainty is the Almahata Sitta meteorite and near-

Earth asteroid 2008 TC$_3$, which impacted the atmosphere in October, 2008. Fragments of the asteroid "rained" down over the Sudan, were collected, and then identified as a ureilite (Jenniskens et al. 2009).

Apart from a compositional link, a dynamical pathway to deliver the meteorites from the asteroid to the Earth is needed. While challenges remain, we can now attempt to dynamically link an asteroid with a meteorite group since the Yarkovsky effect (Bottke et al. 2006) is known to be able to supply small fragments (<10 km) to meteorite-supplying resonances across vast regions of the belt. Bottke et al. (2010) argues that the Yarkovsky effect allows asteroid families far from meteorite-supplying resonances to dominate the meteorite flux. This allows all asteroids in the main belt to contribute meteorites to our terrestrial collection.

A few strong cases can be made for linkages between observed asteroids and meteorite types, such as those discussed in Asteroids III: (4) Vesta, (6) Hebe, and (3103) Eger. Asteroid (4) Vesta (a=2.36 AU) has been known since the 1970s to have visible (McCord et al. 1970) and near-infrared (Larson et al. 1975) spectral properties similar to HEDs. This spectral similarity was confirmed by the Dawn mission and reinforced by bulk elemental composition of Vesta's surface, which are also similar to HEDs (Reddy et al. 2012; De Sanctis et al. 2012; Prettyman et al. 2012). Though Vesta's location in the asteroid belt makes it difficult to produce a significant flux of fragments to the 3:1 and $\nu_6$ meteorite-supplying resonances (Wetherhill 1987), numerous V-type members of the Vesta family (called Vestoids) populate the inner main belt between meteorite-supplying resonances (a=2.22-2.49 AU) (Binzel and Xu 1993) and could serve as potential source bodies. Recently a number of V-types have also been identified in the outer asteroid belt (Lazarro et al. 2000; Hardersen et al. 2004). Although it is very difficult to dynamically move a body from the inner belt to the outer belt, the presence of a number of eucrites with distinctly different oxygen isotopic ratios than most HEDs implies that there could be multiple eucrite parent bodies (McSween et al. 2011), some of which could reside in the outer belt. The recovery of Bunburra Rockhole meteorite (anomalous eucrite) with an Aten-type orbit, delivered to Earth via $\nu_6$ resonance from inner main belt, serves as an additional evidence for multiple basaltic parent bodies (Bland et al., 2009).

S-type asteroid (6) Hebe (a=2.43 AU) was postulated as the parent body of the H chondrites and IIE iron meteorites based on its spectral parameters, large diameter (~200 km), and location near the 3:1 and $\nu_6$ resonances (Gaffey and Gilbert 1998). Recent modeling by Bottke et al. (2010) has confirmed that these resonances are probable source regions of the H chondrites. In addition, asteroid (695) Bella, located near the 3:1 resonance, has been interpreted as having a mineralogy similar to H chondrites (Fieber-Beyer et al. 2012).

E-type near-Earth asteroid (3103) Eger (a=1.41 AU) has been linked with the aubrite meteorites based on its high albedo, flat reflectance spectrum, and absorption feature at ~0.5 µm (Gaffey et al. 1992; Burbine et al. 2002). In the last decade a number of additional E-type asteroids have been observed to have spectral properties consistent with aubrites (Kelley and Gaffey 2002; Clark et al. 2004; Fornasier et al. 2008). Many of these bodies, such as (434) Hungaria, are found among the Hungarias (a=1.78-2.06 AU) (McEarchen et al. 2010). The relatively old median cosmic ray exposure age of aubrites, compared to other meteorite groups, is consistent with being derived from bodies in the Hungaria region (Cuk et al. 2012). Spectra of E-type asteroid (2867) Šteins (a=2.36 AU), the target asteroid of the Rosetta mission, are also consistent with an aubrite composition (e.g. Markus et al. 2013).

Since the publication of Asteroids III, much of the focus on asteroid-meteorite connections has been on the L and LL chondrites. The (1272) Gefion family (a=2.70-2.82 AU) has been proposed by Bottke et al. (2005) and Nesvorný et al. (2009) to be the source of the L chondrites. The Gefion family is composed primarily of S-type objects (Bus 1999) and has an estimated age that is consistent with the 470 Ma shock event (Korochantseva et al. 2007) found in L chondrites. Estimated cosmic ray exposure ages of fragments of the Gefion family, travelling through the 5:2 resonance, are consistent with L chondrites (Nesvorný et al. (2009). A few observed Gefion family members have interpreted mineralogies roughly consistent with L chondrites (Blagen 2012; Roberts and Gaffey 2014); however, this is not the case for all observed members (Roberts and Gaffey 2014).

A relatively large percentage of observed near-Earth asteroids have spectral features and interpreted mineralogies similar to LL chondrites. In fact, approximately two-thirds of S- and Q-type NEAs studied by Vernazza et al. (2008) have olivine/(olivine+pyroxene) ratios similar to LL chondrites. The same is true for members of the (8) Flora family (a=2.16-2.40 AU), which is located near the $\nu_6$ resonance. However, Gaffey (1984) argued that (8) Flora has a non-chondritic mineralogy due to interpreted mineralogical variations that are not consistent with an undifferentiated surface. The LL chondrite mineralogy of at least one S-type asteroid, (25143) Itokawa, has been confirmed by the Hayabusa sample return mission (Nakamura et al. 2011). The Chelyabinsk, Russia meteorite is also an LL chondrite, though the exact parent body is still debated. Reddy et al. (2014) suggested that it might be a fragment of the Flora asteroid family based on mineralogical analysis, while de la Fuente Marcos and de la Fuente Marcos (2014) suggested that the most likely source body was 2011 $EO_{40}$, although its composition is unknown.

Another recent link has been made between the CM chondrites and Ch-type asteroids (13) Egeria (a=2.58 AU) and (19) Fortuna (a=2.44 AU) (Fornasier et al. 2014). Both objects have a distinctive spectral feature at 0.7 μm that is only found in the spectra of CM meteorites. Vilas and Gaffey (1989) study showed that spectra of dark asteroids showed diagnostic carbonaceous chondrite-like absorption bands. Burbine (1998) argued that the location of these bodies near meteorite supplying resonances would be extremely favorable to supplying meteorites to Earth from these bodies, since the relatively fragile CM chondrites would not be expected to be able to survive long in space. However, a number of other asteroids have been classified as Ch-types (Bus and Binzel 2002b; DeMeo et al. 2009) and could also be potential CM chondrite parent bodies. Fieber-Beyer et al. (2012) identified (1358) Gaika, which is also located near the 3:1 resonance, as having an interpreted mineralogy similar to CM chondrites.

## 9. OUTSTANDING QUESTIONS AND FUTURE DIRECTIONS

Despite the advances in asteroid spectroscopy over the last decade, including our ability to constrain the mineralogies of asteroids from which ~90% of meteorites in our terrestrial collection come from, several challenges remain ahead including our inability to identify parent bodies of iron meteorites.

- With the addition of each new calibration for deriving asteroid mineralogy, our ability to characterize the mineralogies of asteroid surfaces improves significantly. However, because different formulas work better for different assemblages, caution must be used when applying these calibrations to any asteroid spectrum. Additional laboratory characterization of samples is necessary before a complete working inventory of calibrations for deriving mineralogies from VIS/NIR spectra can be developed.

- Currently we do not have a reliable method to constrain the mineralogy of most carbonaceous C-type asteroids, M-type asteroids, and primitive achondrite type asteroids. Our inability to treat complex mineralogies is the biggest challenge as we move ahead into the next decade of small body characterization. Some recent advances have been made in showing that most carbonaceous chondrites (widely believed to derive from C-type asteroids) do exhibit absorption features that are diagnostic of their mafic silicates and calcium-aluminum inclusions (Cloutis et al. 2010, 2012).

- Future work is needed to better establish relationships between MGM derived bands and their corresponding mineralogical interpretations, particularly for compositions relevant to asteroids. Such calibrations exist for lunar and martian compositions, but are less common for asteroid studies.

- Future work in radiative transfer modeling of regolith systems should continue to examine the importance of assumptions about particle sizes, shapes, compositions, and particle size distributions. For example, the treatment of complex particles such as agglutinates and impact glasses remains largely unexplored, but could be an important component of future modeling efforts. The planned sample return missions OSIRIS-REx and Hayabusa II will ultimately provide ground truth to improve upon the accuracy of radiative transfer models.

- The availability (or lack) of optical constants for minerals relevant to asteroidal compositions can be a limitation of radiative transfer models. Radiative transfer methods are most effective at constraining compositions for S- and V-type asteroids (and their meteoritic counterparts) due to the wide availability of optical constants for Fe-bearing silicates (e.g. Lucey, 1998; Denevi et al., 2007; Trang et al., 2013). These models are not as successful in treating Fe-poor systems such as Mercury (Lucey and Riner, 2011). Of potential importance to the compositional interpretation of asteroids such as M-types is the need for optical constants of opaque minerals beyond those measured for iron metal (Cahill et al. 2012). There is also a need for improved optical constants at UV wavelengths (particularly for silicates) where little compositional modeling of asteroids has been conducted.

*Acknowledgments.* The authors are grateful for the comments of the editor (F. DeMeo) and two reviewers (P.S. Hardersen, E.A. Cloutis) whose comments helped to improve this chapter. V.R. would like to thank J.A. Sanchez for valuable comets on the effects of temperature and phase angle on near-IR spectra and L. Le Corre for Vesta maps used in this chapter. N.A.M. is grateful to S. Lawrence in drafting the Hapke modeling section. Support for V.R. to carry out this effort came from NASA Planetary Geology and Geophysics Grants NNX14AN35G (PI:Reddy), NNX14AN05G (PI:Gaffey) and NASA Planetary Mission Data Analysis Program Grant NNX14AN16G (PI: Le Corre). V.R., C.A.T., N.A.M. and T.H.B. were visiting astronomers at the Infrared Telescope Facility, operated by the University of Hawaiʻi under contract to the National Aeronautics and Space Administration.

**Table 1. Calibrations for correcting non-compositional effects**

Table 1. Calibrations for correcting temperature effects

| Effect | Analogue | Operation | Equation | Equation No. |
|---|---|---|---|---|
| Temperature | Howardites and Eucrites[a] | Add to Band I Center | $\Delta BI(\mu m) = 0.01656 - 0.0000552 \times T(K)$ | 2 |
| | | Add to Band II Center | $\Delta BII(\mu m) = 0.05067 - 0.00017 \times T(K)$ | 3 |
| | Diogenites[a] | Add to Band I Center | $\Delta BI(\mu m) = 1.7 \times 10^{-9} \times T^3(K) - 1.26 \times 10^{-6} \times T^2(K) + 2.66 \times 10^{-4} \times T(K) - 0.0124$ | 4 |
| | | Add to Band II Center | $\Delta BII(\mu m) = 0.038544 - 0.000128 \times T(K)$ | 5 |
| | Ordinary Chondrites[b] | Add to Band II Center | $\Delta BII(\mu m) = 0.06 - 0.0002 \times T(K)$ | 6 |
| | | Add to Band Area Ratio | $\Delta BAR = 0.00075 \times T(K) - 0.23$ | 7 |
| | Olivine-rich A-type Asteroids[b] | Add to Band I Center | $\Delta BI(\mu m) = -(1.18 \times 10^{-7})T^2 + (2.15 \times 10^{-5})T + 0.004$ | 8 |
| | | Add to Band I Depth | $BI_{depth.corr} = 0.0057 \times T(K) - 1.71$ | 9 |

[a]Reddy et al. (2012), [b]Sanchez et al. (2012)

**Table 2. Calibrations for various meteorite mineralogies and their asteroid analogues**

| Asteroid type | Analogue | Comp. range | Equation | Error |
|---|---|---|---|---|
| V-types | HEDs[a] | Fs23-56; Wo2-14 | Fs = 102.3 * BIC - 913.82 | ± 3 mol% |
| | | | Fs = 205.9 * BIIC - 364.3 | ± 3 mol% |
| | | | Wo = 396.1 * BIC - 360.55 | ± 1 mol% |
| | | | Wo = 79.91 * BIIC - 148.3 | ± 1 mol% |
| S-types | Ordinary Chondrites[b] | Fa17-31 | Fa = -1284.9 * (BIC)$^2$ = 2656.5 * (BIC) - 1342.2 | ± 1.3 mol% |
| | | Fs14-26; Wo1-3 | Fs = -879.1 * (BIC)$^2$ + 1824.9 * (BIC) - 921.7 | ± 1.4 mol% |
| | Lodranites/ Acapulcoites[c] | | ol/(ol+px) = -0.242 * BAR + 0.728 | ±0.03 |
| | | | px/(ol+px) = 0.417 * BAR + 0.052 | |
| A-types | R chondrites[d] | Fo10-90 | Fo = -1946.6 * (BIC) - 2139.4 | ± 5 mol% |
| | | | ol/ol+px = -11.27 * BAR$^2$ + 0.3012 * BAR + 0.956 | |

[a]Burbine et al. (2007), [b]Dunn et al. (2010b), [c]Cloutis et al. (1986), and [d]Reddy et al. (2011)

**Figure Captions**

Figure 1. Laboratory near-IR spectra of Dhurmsala meteorite (LL6 ordinary chondrite) at different phase angles α = 13° (solid line), 30° (short dashed line), 60° (long dashed line), 90° (dot-short dashed line) and 120° (dot-long dashed line) from Sanchez (2012). The figure shows spectra that are not normalized or offset from each other. As the phase angle increases the overall reflectance decreases and the two arrows show the dramatic decrease in reflectance at the two the reflectance peaks of the spectra. The reflectance in the shorter wavelength end (blue region) of the spectrum decreases faster than the longer wavelength end (red) of the spectra with increasing phase angle creating the observed red slope (phase reddening).

Figure 2. Laboratory-measured spectra of olivine (Fa35) in the brachinite EET 994012 (Burbine et al. 2007) and pyroxene (Fs53Wo14) in the eucrite Bouvante (Burbine et al. 2001). Area under the dashed lines is the Band area. The Band area ratio is the area of Band II/area of band I.

Figure 3. Flowchart showing a decision making tree for mineralogical characterization of A-, S-, and V-type asteroids using equations from Reddy et al. (2011), Dunn et al. (2010), and Burbine et al. (2007).

Figure 4. Modified Gaussian model of asteroid 25143 Itokawa. From bottom to top: asteroid data with error bars (black), best fit model (grey), model continuum (dashed), six Gaussian bands used in the model, and residuals from the model fit. The Gaussian band centered around 0.9 microns (thick black) is attributed to pyroxene (pyx), while the band around 1.2 microns (thick grey) is attributed to olivine (ol). The ratio of these bands strength's provides a means for comparison to equivalent models of OC meteorite spectra. This comparison correctly predicts an LL-type composition for Itokawa (Binzel et al. 2001).

Figure 5. Shkuratov model of asteroid 25143 Itokawa. From bottom to top: asteroid data with error bars (black), best fit model (grey), spectra of mineral endmembers olivine (ol) and low-Ca pyroxene (pyx) used in the model, and residuals from the model fit. This model, based on Vernazza et al. (2008), assumes 50% porosity on the asteroid's surface and only fits the data between 0.55 and 2 microns due to inaccuracy in the treatment of space weathering at longer wavelengths (see Brunetto et al. 2006 for details). This model suggests a composition of ol/(ol + low-Ca pyx) = 76%, consistent with the known LL-chondrite composition of Itokawa.

Figure 6. (A) Plot showing olivine iron content (mol.% fayalite) on the Y-axis and pyroxene iron content (mol.% ferrosilite) on the X-axis from laboratory measurements of ordinary chondrites (H, L and LL). Fayalite and ferrosilite values from Itokawa samples (Nakamura et al. 2011) and spectrally derived values using Binzel et al. (2001) IRTF telescopic spectra data are also shown. Spectrally derived values were calculated using equations from Dunn et al. (2010). The difference between the olivine and pyroxene iron content between the two datasets is less than 1 mol.% and plot in the LL chondrite zone

as predicted by Binzel et al. (2001). Similarly we plot the Fayalite and Ferrosilite values for (433) Eros from Murchie and Pieters (2003) (ground-based spectral data) and Izenberg et al. (2003) (NEAR NIS spectral data) in Figure X (B). The Fs and Fa values for Eros are identical and fall in the L chondrite region of the plot. The figure is adapted from Nakamura et al. (2011).

Figure 7. (A) HST map of Vesta in 0.673-µm filter projected in the Thomas et al. (1997a) coordinate system based on observations from 1994, 1996 and 2007 oppositions at a resolution of ~50 km/pixel. (B) Dawn FC map of Vesta in 0.75-micron filter from Rotational Characterization 1 (RC1) phase at a resolution of 9.06 km/pixel with prime meridian similar to Thomas et al. (1997a) coordinate system. The maps show corresponding bright features and dark features identified by Binzel et al. (1997) and Li et al. (2010). The figure is reproduced from Reddy et al. (2013).

**Figures**

Figure 1.

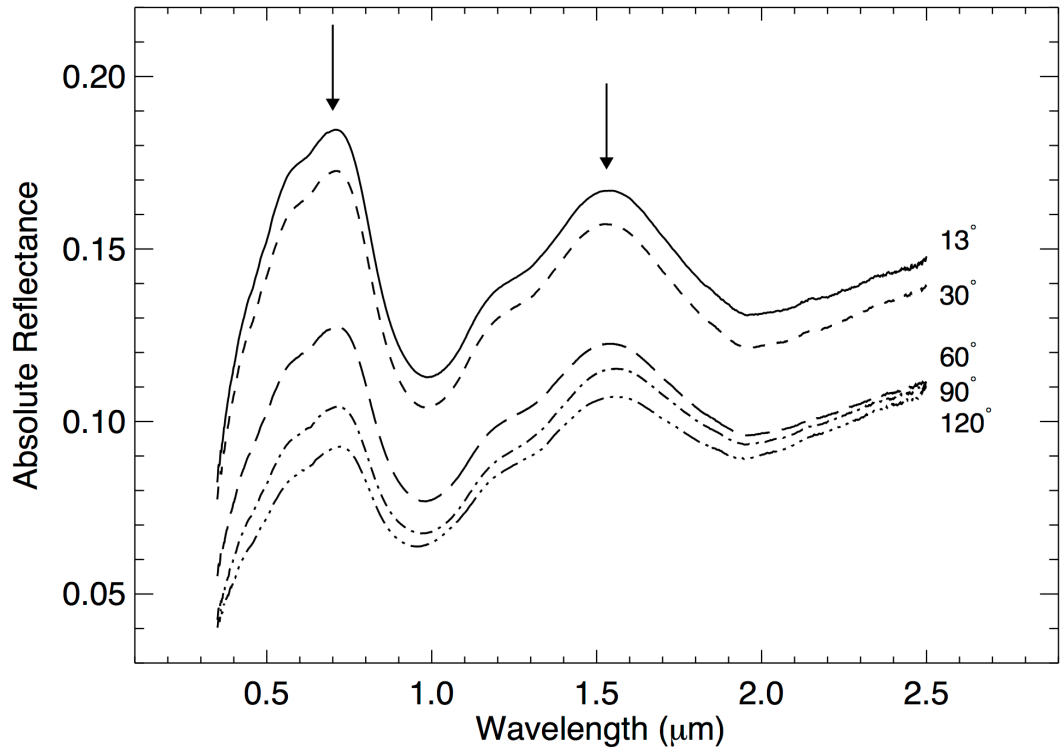

Figure 2.

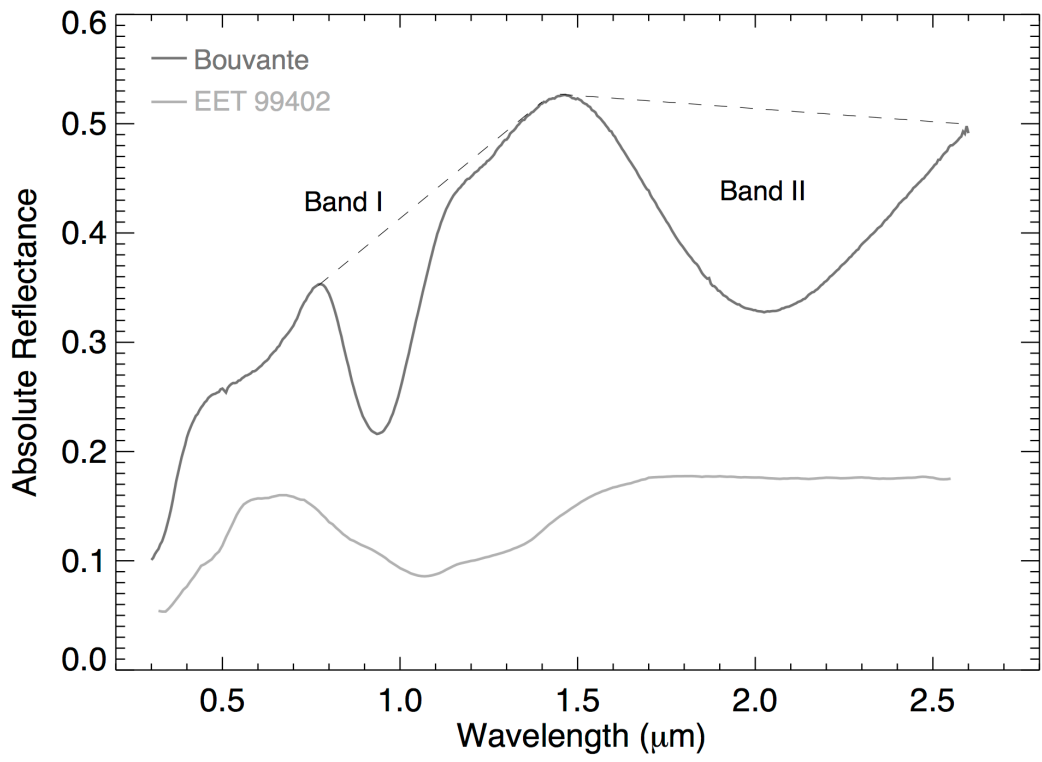

Figure 3

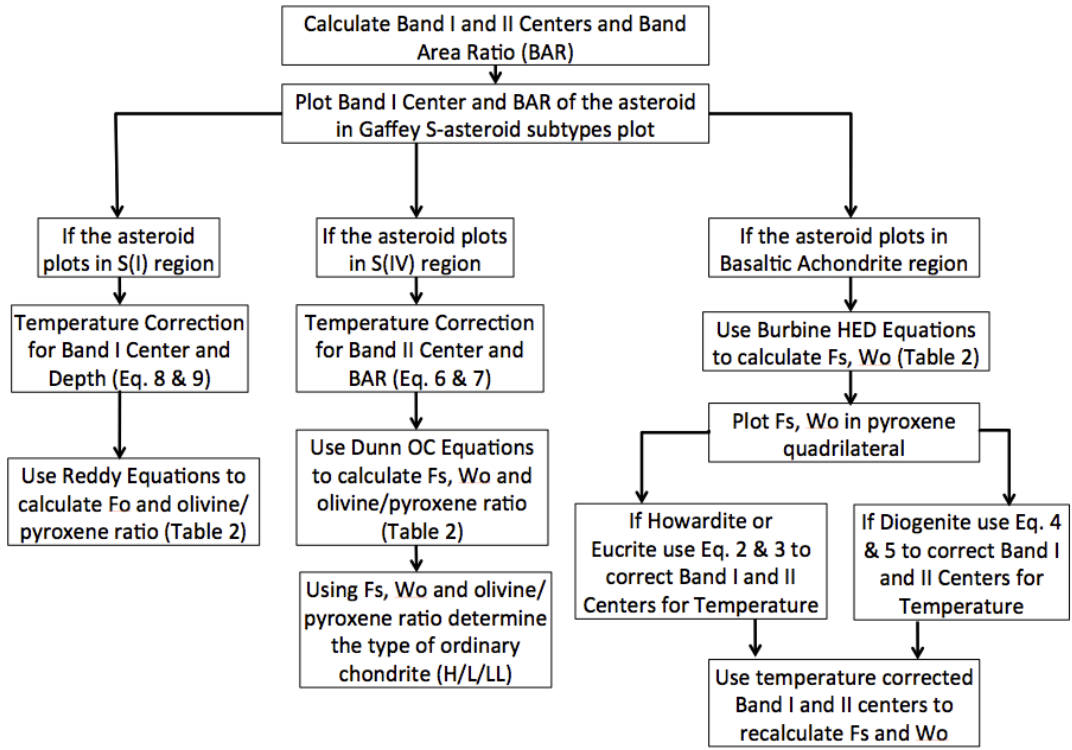

Figure 4.

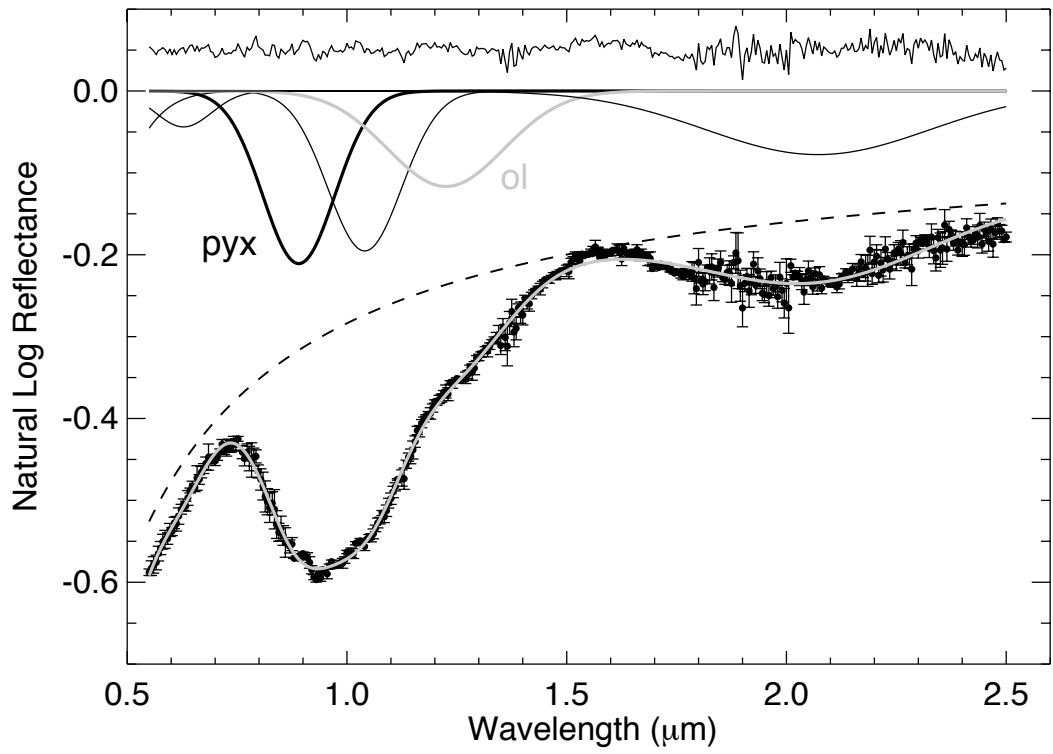

Figure 5.

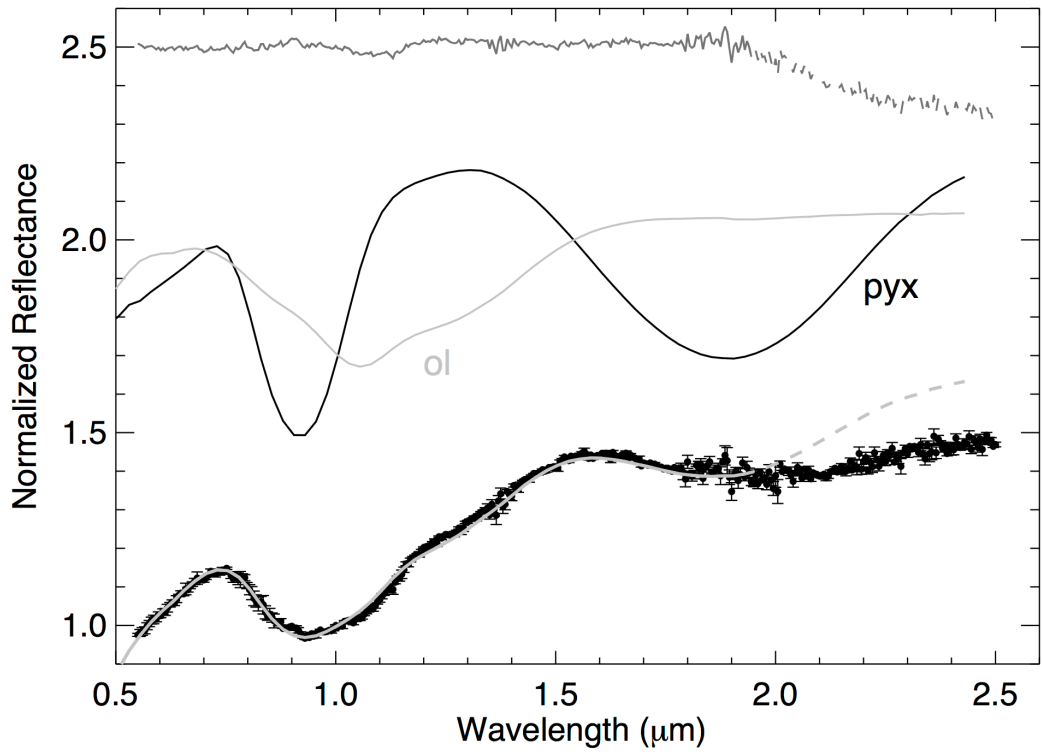

Figure 6.

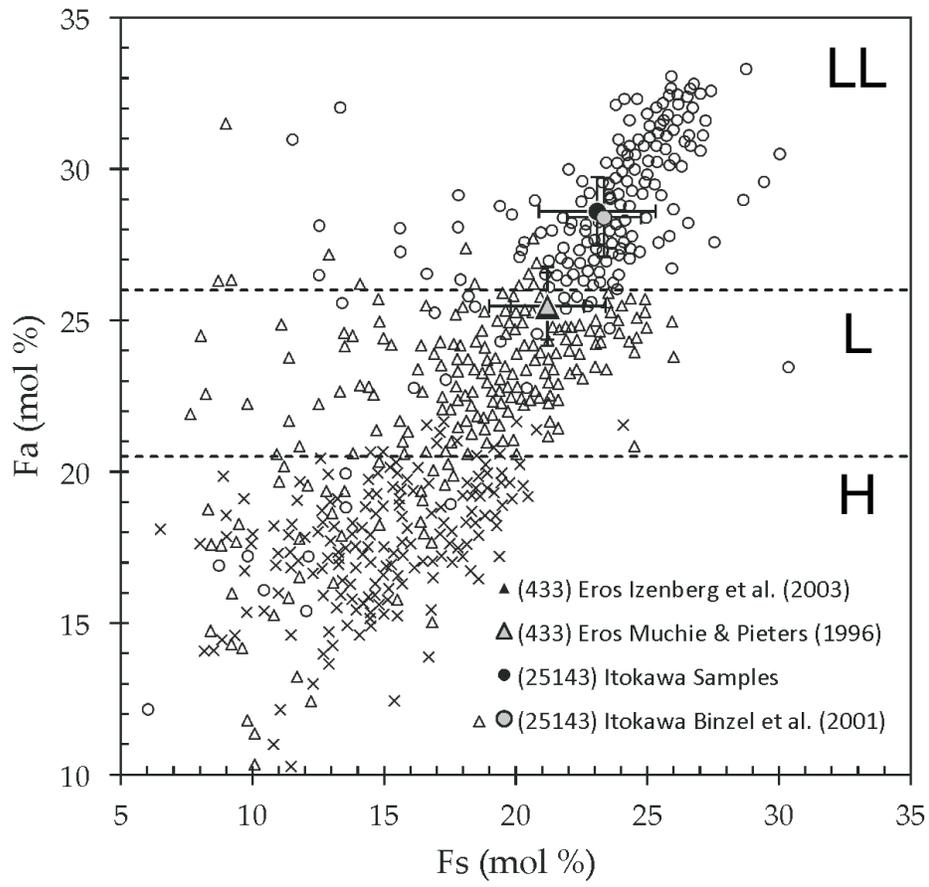

Figure 7.

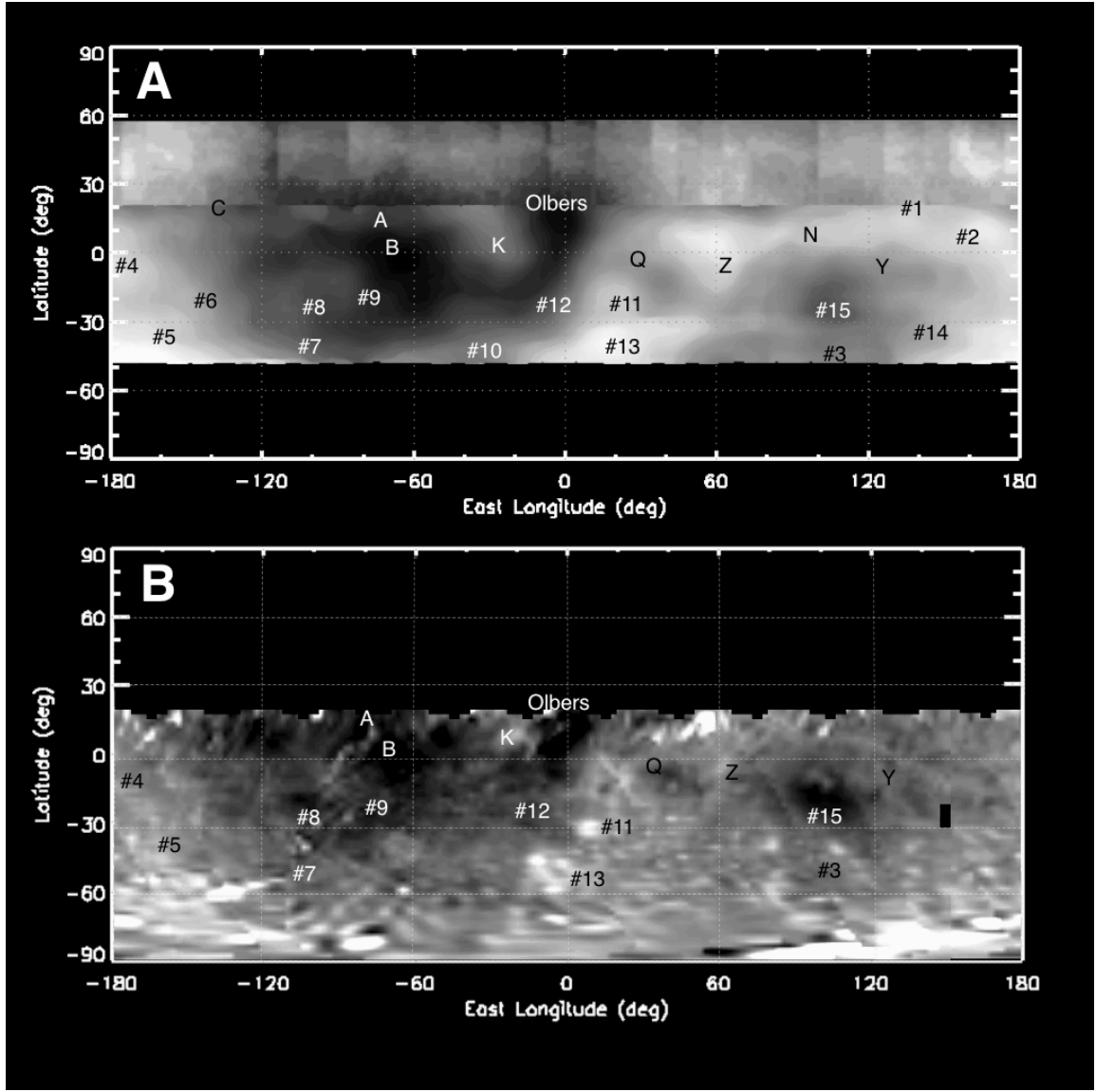